\documentclass{PHYEAUTH}
\usepackage{graphicx}
\usepackage{amsmath}
\usepackage{amssymb}
\begin{document}

\begin{frontmatter}

\title{Electron magnetotransport in GaAs/AlGaAs superlattices with
  weak and strong inter-well coupling}
\author[address1]{L.\ Smr\v{c}ka},
\author[address1]{N.\ A.\ Goncharuk\thanksref{thank1}},
\author[address1]{P.\ Svoboda},
\author[address1]{P.\ Va\v{s}ek},
\author[address1,address2]{Yu.\ Krupko}
and
\author[address3]{W.\ Wegscheider}
\address[address1]{Institute of Physics, Academy of Sciences of the
Czech Republic, v.v.i., Cukrovarnick\'a 10, 162 53 Prague 6, Czech Republic}
\address[address2]{Grenoble High Magnetic Field Laboratory, B.P. 166,
F-38042 Grenoble Cedex 9, France}
\address[address3]{Universit\"at Regensburg, Universit\"atstrasse 31,
D-93040 Regensburg, Germany}
\thanks[thank1]{
Corresponding author.
E-mail address: gonchar@fzu.cz}
\begin{abstract}
We report on magnetotransport measurements in 
two MBE-grown GaAs/AlGaAs superlattices 
formed by wide and narrow quantum wells and thin Si-doped barriers 
subject to tilted magnetic fields. 
It has been shown that illumination of the strongly coupled superlattice
with narrow wells leads to reduction of its dimensionality from the 3D
to 2D. The illumination-induced transition is revealed by remarkable change of
magnetoresistance curves as compared to those measured before
illumination. 
The experimental data along with tight-binding model calculations
indicate that the illumination not only enhances the electron
concentration but also suppresses the electron tunneling
through the barriers.  
\end{abstract}
\end{frontmatter}
%
\section{Introduction}
A semiconductor superlattice (SL) is a periodic 
heterostructure made
of repeated layers of two host materials which are structurally
similar but possess different electrical properties \cite{Esaki}. 
The periodicity of the one--dimensional SL potential along the layer
growth direction $z$ with the period 
$d_z = d_w+d_b$ [composed of the quantum well (QW) width $d_w$ 
and the barrier thickness $d_b$] gives
rise to the energy miniband structure within the conduction band of
the host material. 
The SL thus provides a unique artificial system 
where the electron transport through
the minibands can be controlled by selecting 
the parameters $d_w$, $d_b$ and the effective barrier height.  
 
Wider QWs and/or thicker barriers characterize a weakly coupled SL
with small or even negligible overlap of electron wave functions from
adjacent wells. The corresponding electron system tends to be
two-dimensional (2D) with an open anisotropic Fermi surface (FS) of
cylindrical shape and SL minibands 
reduced to discrete energy levels. 
The wave function overlap grows upon decreasing the thickness
of the layers that form the SL. It introduces a broadening of SL
minibands and the electron system becomes effectively
three-dimensional (3D). In the limiting case of strong inter-well
coupling the FS transforms to a set of closed spheroids. 

We use magnetotransport measurements in strong magnetic fields tilted 
with respect to the QW planes to detect the SL dimensionality 
by reconstructing the FS in $k$-space \cite{Onsager}. 
It has been experimentally proved \cite{Jaschinski} and 
theoretically supported (see e.g. Ref.~\cite{Goncharuk}) 
that a large in-plane magnetic 
field component $B_y$ suppresses the inter-well coupling 
and makes thus a 3D$\rightarrow$2D transition possible. In this paper 
we demonstrate another experimental possibility how to induce the
3D$\rightarrow$2D transition
by illuminating a strongly coupled SL.       

\section{Experiments}
Two different MBE-grown SL structures have been employed in this
study. Both comprised 30 periods with $d_z = 24$~nm (sample {\it A}) or 
$d_z = 8.5$~nm (sample {\it B}), each of them therefore consisted of 29 GaAs
QWs separated by Si-doped $\rm Al_{0.3}Ga_{0.7}As$ barriers. The
parameters of the weakly coupled structure {\it A} were  $d_{w}=19$~nm
and $d_{b}=5$~nm, those of the strongly coupled structure {\it B} were 
$d_{w}=4.5$~nm and $d_{b}=4$~nm. All the barriers were composed of
central Si-doped parts of thickness 3.7~nm and 2.7~nm for structures
{\it A} and {\it B}, respectively, surrounded by undoped spacer layers
0.65~nm thick.  

Several samples in the Hall bar geometry have been patterned 
by means of the optical lithography, equipped with evaporated AuGeNi 
contacts and adjusted to ceramic chip carriers. 
The chip carrier has been attached to a rotatable plate in a $^3$He insert. 
The plate allowed to rotate samples to any angle between the 
perpendicular ($\varphi=0^{\circ}$, $B=B_z$) and in-plane
($\varphi=90^{\circ}$, $B=B_y$) magnetic field orientations.
Both the longitudinal ($\rho_{xx}$) and Hall ($\rho_{xy}$) 
resistances were measured at $T$ = 0.4 K in
magnetic fields up to 28 T (sample {\it A}) and 
$T$ = 0.3 K, up to 13 T (sample {\it B}). 
The standard low-frequency ($f$ = 13 Hz) lock-in technique has been 
employed for the measurement.

A red LED was installed close to the sample {\it B} to
facilitate its "in situ" illumination in liquid $^3$He. Short current pulses 
($\approx 100$~msec) were successively applied to the 
LED while monitoring the decrease of the zero field resistance 
of the sample. A few such pulses were sufficient to reach 
the saturation of the resistance. All the curves reported below 
for the illuminated sample {\it B} refer to this saturated state.

\section{Results and  discussion}
\begin{figure}[!b]
\begin{center}
\includegraphics[width=0.683\linewidth]{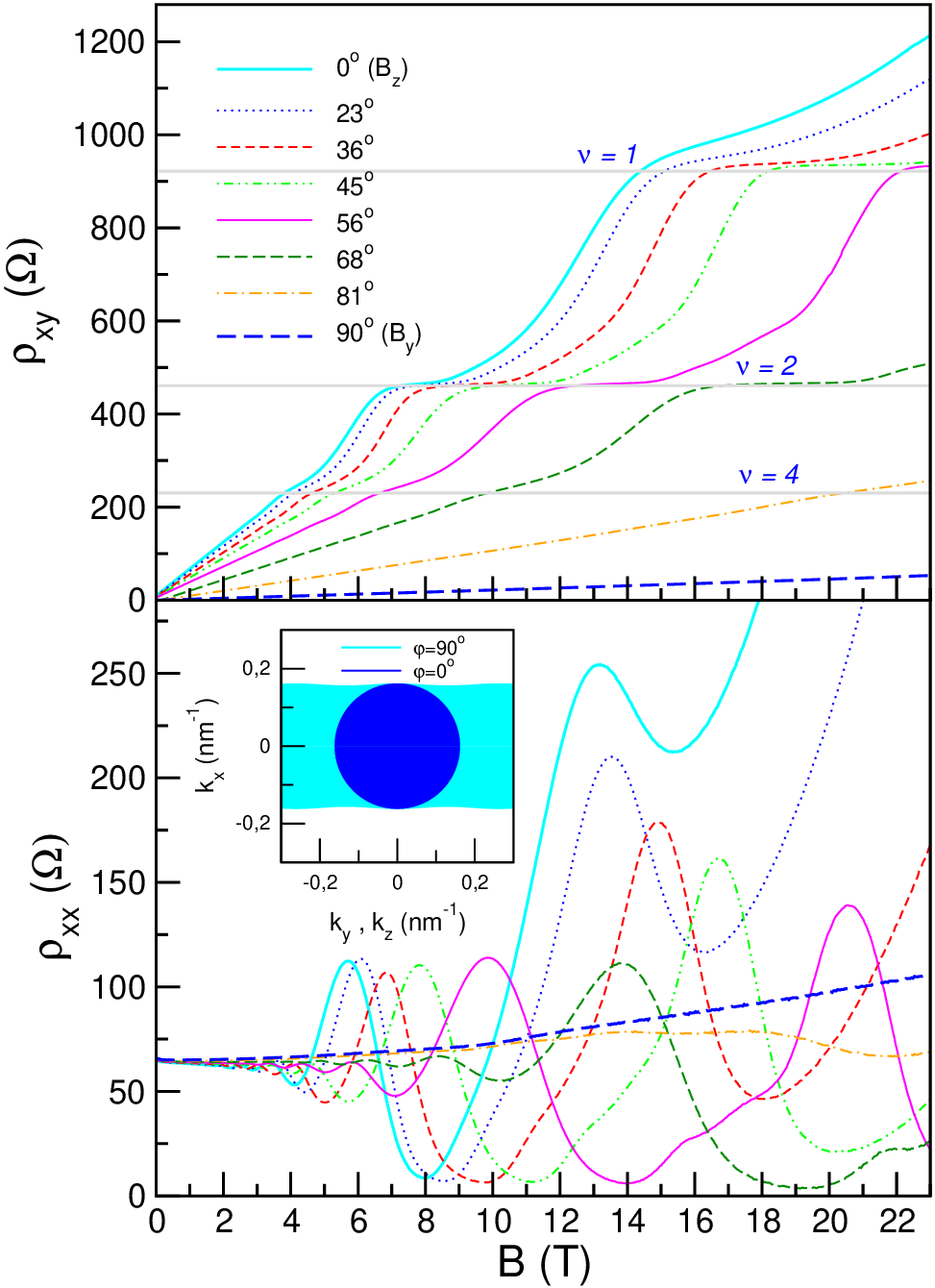}
\caption{\label{fig1} 
Transversal and longitudinal magnetoresistance curves of 
the weakly coupled SL {\it A} 
at various tilt angles $\varphi$. The insert shows 
cross-sections of the FS for two limited configurations of the applied
magnetic field, normal and parallel to the sample plane.} 
\end{center}
\end{figure}

Experimental curves for a set of tilt angles $\varphi$ collected on the 
sample {\it A} are displayed in Fig.~\ref{fig1}. 
\begin{figure}[!t]
\begin{center}
\begin{tabular}{c}
\hbox{
\includegraphics[width=0.683\linewidth]{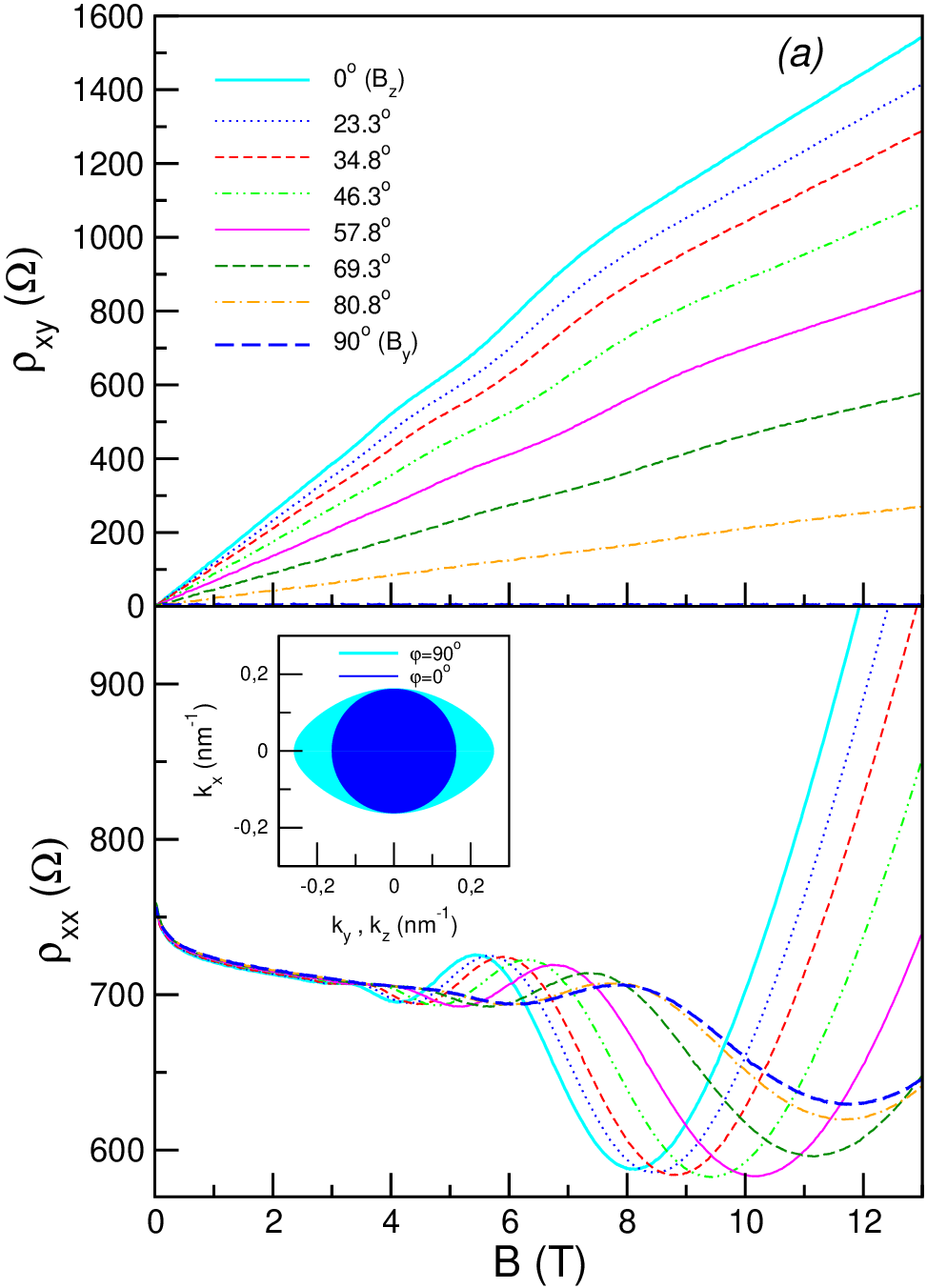}} \\
\vbox{
\includegraphics[width=0.683\linewidth]{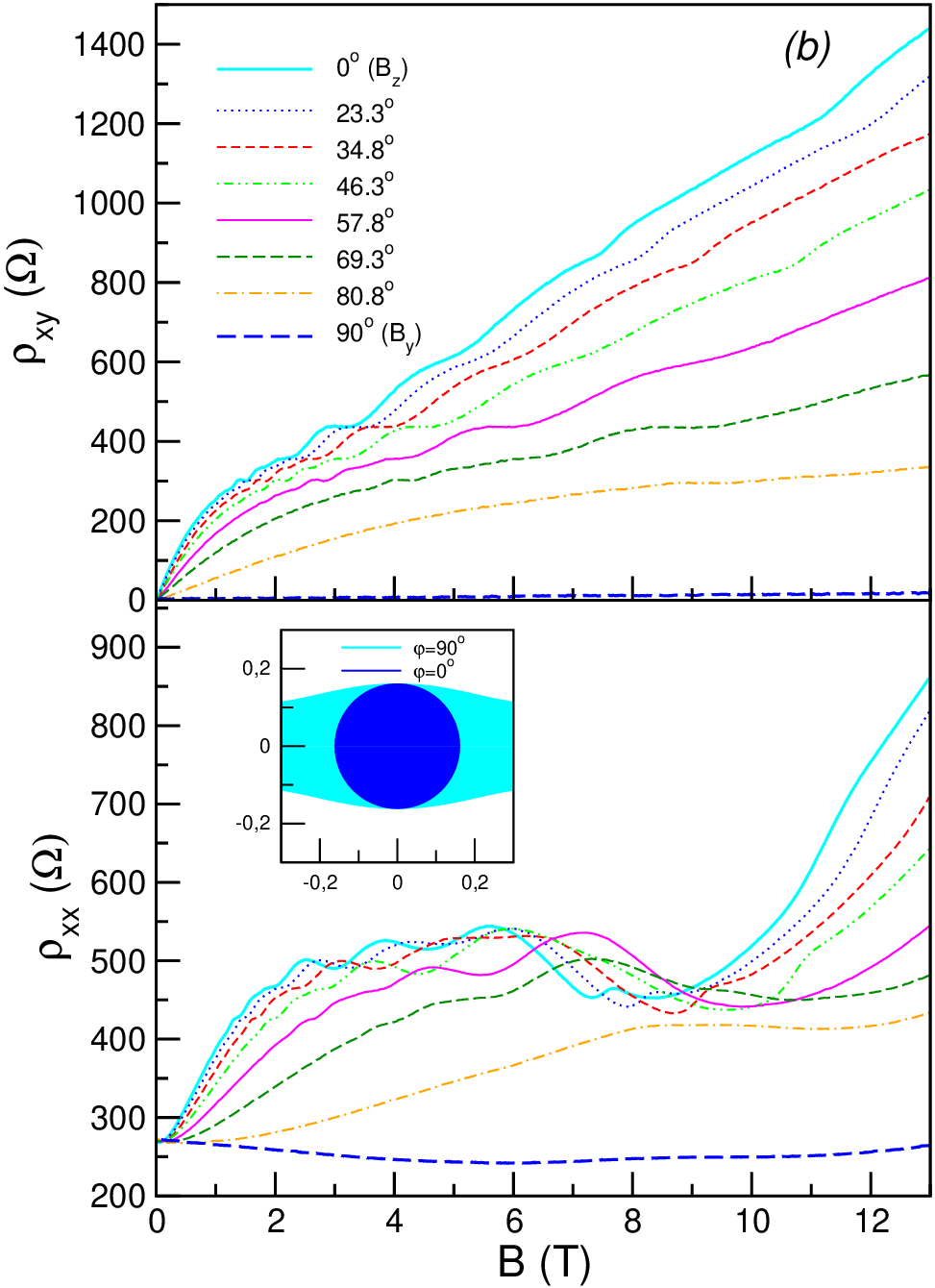}}
\end{tabular}
\caption{\label{fig2} 
Transversal and longitudinal magnetoresistance curves
measured at various tilt angles $\varphi$ 
on the SL {\it B} ``in dark'' {\it (a)} and
after illumination by light pulses {\it (b)}. 
}
\end{center}
\end{figure}
From the occurence of the quantum Hall effect we may conclude that the 
SL {\it A} represents a nearly perfect 2D system with negligible
coupling of the 2D electron layers confined in the QWs. 
From the positions of the quantum Hall electron plateaux 
($\rho_{xy}=h/ije^2$, where $i,j$ are integer numbers and $j$ is the 
amount of QWs in the SL) we estimated
that 28 QWs are occupied by electrons, 
which is very close to the nominal value 29. 
The electron concentration per layer 
$N_{H}=3.9\times 10^{11}$~cm$^{-2}$ extracted from the linear part of 
the Hall magnetoresistance is in a good agreement with the
value $N_{SdH}=4.0\times 10^{11}$~cm$^{-2}$ obtained from the period of
SdH oscillations at $\varphi = 0^{\circ}$. 
Corresponding electron mobility per layer is $\mu_H =
8.7\times10^3$~cm$^2$/V$\cdot$sec.
Oscillations of $\rho_{xx}$ are shifted to higher fields with tilting
the angle, satisfying the cosine law. 

The FS of the sample {\it A} has a form of a smoothly corrugated
(almost flat) cylinder 
(see the inset in Fig.~\ref{fig1}) that was expected for the slightly
coupled SL. It is difficult to detect the positions of ``belly'' (at
the center of the Brillouin zone, $k_z=0$) and
``neck'' (at the Brillouin zone boundry, $k_z=\pm\pi/d_z$) 
extremal cross-sections of the FS, as their areas are nearly the same.
 
Magnetotransport data measured on the sample {\it B} before and after
illumination are presented in Fig.~\ref{fig2}.
Magnetoresistance curves of the SL ``in dark'' exhibit 
strong deviations from the 2D behavior 
seen in the sample {\it A}.
The Hall resistance grows almost linearly with increasing field. 
SdH oscillations can be seen for all slopes of the magnetic 
field even in the strictly in-plane orientation.  
These oscillations are found to be periodic in 1/B with a single period. 
Thus, the FS of the SL {\it B} ``in dark'' is formed by an ellipsoid
[see the inset in Fig.~\ref{fig2} {\it (a)}] that
confirms the 3D nature of the strongly coupled SL sample.

Illumination of the sample {\it B} dramatically changed the
magnetoresistance curves. 
Hall plateaux appeared on
$\rho_{xy}$ curves. Oscillations of $\rho_{xx}$ acquired more
complicated forms and exhibit two distinct periods originated from
``belly'' and ``neck'' orbits. 
Oscillations weaken with increasing $\varphi$ and disappear completely
at $\varphi=90^{\circ}$. It means that the SL dimensionality has been
reduced after illumination. 
The FS constructed from experimental periods of ``belly'' and ``neck'' 
SdH oscillations is another evidence. 
After illumination the FS of the SL {\it B} 
has changed its shape to a corrugated cylinder 
[see the inset in Fig.~\ref{fig2} {\it (b)}] characteristic for a
system composed of weakly-coupled 2D electron layers.
 
As expected, the carrier concentration per
layer was enhanced after illumination from the value 
$N_{H}=1.7\times 10^{11}$~cm$^{-2}$ to 
$N_{H}=2.2\times 10^{11}$~cm$^{-2}$. The electron mobility per layer
increased significantly from 
$\mu_H=1.7\times10^3$~cm$^2$/V$\cdot$sec to 
$\mu_H=3.6\times10^3$~cm$^2$/V$\cdot$sec.
Interpretation of the data in terms of the theory \cite{Goncharuk}
simultaneously leads to a conclusion, that the ground SL miniband
shrinks after illumination and therefore the inter-well coupling
becomes weaker. Such a surprising behavior may be due to the presence 
of the metastable ionized deep  donors within the barriers, 
which effectively increases their heights.

\section*{Acknowledgements}
This work has been supported by the European Commission contract
No. RITA-CT-2003-505474, Ministry of Education of the Czech Republic
Center for Fundamental Research LC510 and Academy of Sciences of the
Czech Republic project KAN400100652.

\end{document}